\documentclass{egpubl}
\usepackage{eurovis2026}
\usepackage{xurl}

\EuroVisShort  

\usepackage[T1]{fontenc}
\usepackage{dfadobe}

\usepackage{cite}  
\BibtexOrBiblatex
\electronicVersion
\PrintedOrElectronic
\ifpdf \usepackage[pdftex]{graphicx} \pdfcompresslevel=9
\else \usepackage[dvips]{graphicx} \fi

\usepackage{egweblnk}

\newcommand{\tool}{AwesomeLit}

\title[\tool{}]%
      {\tool{}: Towards Hypothesis Generation with Agent-Supported Literature Research}

\author[Z. Xie \& Y. Guo \& K. Xu]
{\parbox{\textwidth}{\centering  Z. Xie$^{1}$\orcid{0009-0009-0770-1428}, Y. Guo$^{2}$\orcid{0009-0004-3857-7486}
       and K. Xu$^{1}$\orcid{0000-0003-2242-5440} 
        }
        \\
{\parbox{\textwidth}{\centering $^1$School of Computer Science, University of Nottingham, UK\\
$^2$ School of Intelligence Science and Technology, Peking University, China
}
}
}

\begin{document}

\teaser{
  \includegraphics[width=\linewidth]{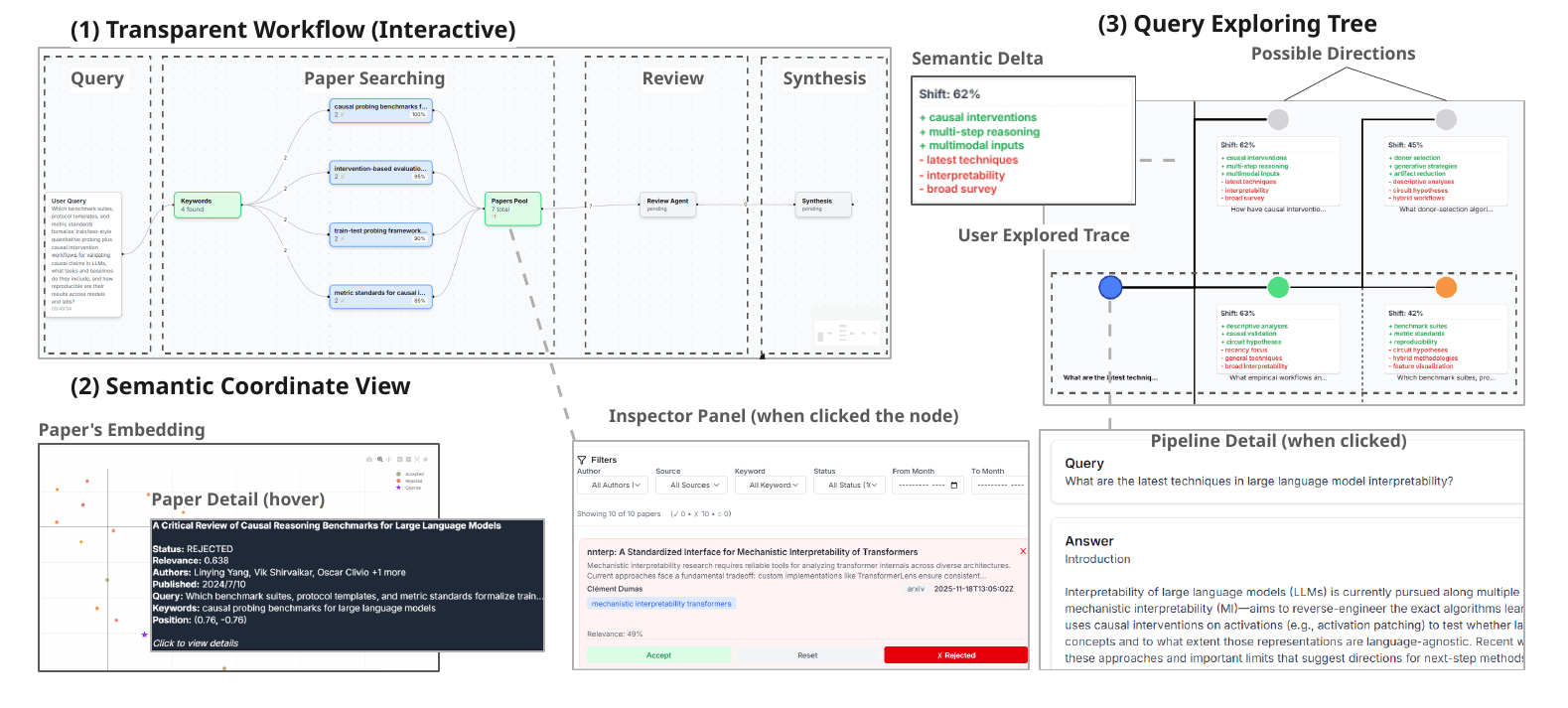}
  \centering
  \caption{Overview of the \tool{}: it visualizes the workflow (1) containing query, paper searching, review and synthesis nodes. A Semantic Similarity View (2) is shown to provide contextual grounding for the relationships between different papers and the shift of queries. Besides a Query Exploring Tree (3) is displayed to externalize user's exploration path.}
  \label{fig:teaser}
}

\maketitle
\begin{abstract}
   There are different goals for literature research, from understanding an unfamiliar topic to generate hypothesis for the next research project. The nature of literature research also varies according to user's familiarity level of the topic. For inexperienced researchers, identifying gaps in the existing literature and generating feasible hypothesis are crucial but challenging. While general ``deep research'' tools can be used, they are not designed for such use case, thus often not effective. In addition, the ``black box" nature and hallucination of Large Language Models (LLMs) often lead to distrust. In this paper, we introduce a human-agent collaborative visualization system \tool{} to address this need. It has several novel features: a transparent user-steerable agentic workflow; a dynamically generated query exploring tree, visualizing the exploration path and provenance; and a semantic similarity view, depicting the relationships between papers. It enables users to transition from general intentions to detailed research topics. Finally, a qualitative study involving several early researchers showed that \tool{} is effective in helping users explore unfamiliar topics, identify promising research directions, and improve confidence in research results.


\end{abstract}  
\section{Introduction}
Literature research is a critical task in both industry and academia. Common industrial examples include market research and horizon scanning. In the academia, literature research is an integral part of understanding latest research development, forming new research ideas, and producing publications such as systematic literature review~\cite{PMB*21}. Literature research can require significant time and resources, especially for fast developing field like generative AI. The advent of Large Language Model (LLM) has introduced new possibilities for assisting this task. General LLM tools, such as the ``deep research'' provided by OpenAI or Gemini, can find relevant information, but are not optimized for literature research thus not always effective. For example, they might not have access the latest publications. Also, many existing tools prioritize automation, which creates a ``black box'' that is difficult to interpret and thus leads to less user trust.

The user's level of expertise can also have a notable impact on the nature of literature research: for less-experienced researchers who are not familiar with the topic, literature research is as much a learning experience as an exploration process. In such scenarios, the goal of literature research can be a moving target: as the understanding improves, the goal is constantly refined and updated. Again, this is not well supported in existing tools. For example, such users have been found to frequently struggle to bridge the gap between their vague initial intents and the specialized vocabulary required for effective retrieval~\cite{SOM24}, making the process both time-consuming and energy-draining to oversight. Also, they are more likely to lack the domain knowledge to verify AI results or assess their confidence levels~\cite{SOM24}, making them more vulnerable to AI hallucinations and hindering the development of critical research skills~\cite{KLKK24}.

Furthermore, existing tools typically lack mechanisms for effective Human-AI collaboration. While some of the tools would display the model's reasoning process, there is no easy way to modify them. Also, as mentioned before novice researchers often iteratively refine their prompts as the literature research progresses. However, most existing systems fail to visualize this evolution~\cite{HRM*24}, preventing the formation of a shared mental model between the user and the AI~\cite{XZX*24}. In addition, to cultivate critical judgment~\cite{CLLY24}, novices need visual evidence—such as semantic distributions—to help interpret uncertainty and incorporate AI rationale into their decision-making process~\cite{RBK25}, but this is often lacking in current tools.

This paper introduces \tool{}, which aims to assist literature research by junior researchers who look for research opportunities in an unfamiliar topic area. This is achieved partly by converting complex interactions with LLM agent into clear and controllable workflows. By combining LLM agent with coordinated visualization techniques, this tool enables the monitoring of the agent reasoning process and allow users to continuously refine their research. The interface coordinates a \textbf{Transparent Workflow} for fine intervention, provides a \textbf{Semantic Similarity View} for multi-dimensional paper screening, and a \textbf{Query Exploring Tree} to manage topic transitions. It organizes divergent research branches to facilitate hypothesis generation, supports a seamless transition from broad exploration to in-depth analysis, and bridges the key gap between AI automation and manual verification.

This paper makes two key contributions. First, derived from a formative study, it characterizes the design requirements for human-AI collaborative for literature research in our specific context, highlighting the needs for trust, steerability, and evolutionary inquiry. Second, it presents a novel system that addresses these needs by providing an intuitive interface for the transparent, controllable, and structured exploration of academic literature. The effectiveness of \tool{} is demonstrated with the results from the qualitative study with seven participants from diverse background.

\section{Related Work \& Backgrounds}
Generative AI has fundamentally shifted scientific workflows from passive search to active discovery. Recent studies highlight the transformative potential of autonomous agents in hypothesis generation~\cite{RS25, PSS*25}. However, ensuring these agents align with human intent remains a challenge. Research in collaborative guidance emphasizes the need for iterative co-creation mechanisms~\cite{FZF*25} and multi-modal steering~\cite{CWG*25, LCM25} to manage ambiguity in conceptual design.

While visual analytics has made strides in interpreting LLM internal representations~\cite{SGSE25} or quantifying subjective metrics~\cite{HCW*25}, current XAI frameworks often focus on model debugging rather than supporting the exploratory logic of novice researchers. Although methods for validating LLM outputs exist—such as visual slice discovery~\cite{YXO*25} or trust-based teaming protocols~\cite{AHIM25}—they are rarely integrated into a unified workflow that simultaneously supports generative exploration and rigorous source verification~\cite{SNW*25}. AwesomeLit addresses this gap by synthesizing these diverse signals—generative planning, visual steering, and semantic evidence—into a cohesive system tailored for early-stage inquiry.

In the literature research domain, several systems have been developed to facilitate semantic exploration and autonomous generation. An et al.~\cite{ANWX24} introduce vitaLITy2, a key example in this transformation by leveraging LLM-based embedding techniques to enable researchers to interactively explore research domains. While effective for visualizing static landscapes, they often lack the generative reasoning capabilities required to formulate novel hypotheses. Commercial platforms like Consensus~\cite{Con24} utilize specialized LLMs to directly extract and summarize evidence-based assertions from peer-reviewed papers. However, they typically operate as "black boxes" providing answers without exposing the intermediate retrieval logic, which hinders users from diagnosing hallucinations or steering the search focus. Recent advancements focus on high-fidelity knowledge synthesis. OpenScholar~\cite{AHS*26} and STORM~\cite{SJK*24} demonstrate state-of-the-art performance in generating long-form surveys through multi-perspective questioning. Similarly, long-horizon agents from OpenAI~\cite{Ope25b} and Google~\cite{Goo25}, as well as academic prototypes like Research Agent~\cite{BJCH25}, can autonomously perform complex search-reasoning loops. While these autonomous agents excel at efficiency and automation, they often sideline the user's need for cognitive engagement. They tend to produce a final report directly, bypassing the crucial iterative sensemaking process that novice researchers need to define their own direction. In contrast, our system prioritizes process transparency and human-in-the-loop steering. It visualize the agent's planning workflow and implement granular checkpoints to allow users to intervene and prune branches, ensuring the final hypothesis is a product of collaborative evolution rather than passive consumption.

\section{Requirement Analysis}
The target users were defined as early-stage researchers who want to transform a vague research idea into a solid hypothesis, typically characterized by limited domain expertise, such as capstone students or junior graduates. To inform our design, a formative study with eight target users (five undergraduates, one master, and two doctors) was conducted to guide the AwesomeLit design. The participants performed literature review tasks using the latest AI agent (Consensus~\cite{Con24}) to find interesting directions and the search method (Google Scholar~\cite{Goo05}) to  verify the results. Each session lasted approximately 75 minutes, comprising a 30 minute literature review task and a 45 minute interview. During the task,they were asked to identify a novel research direction in 'Visualization for AI' and produce a brief proposal outline at the end.

Observations revealed significant friction in their workflow, which begins with multi-level metadata filtering like timeliness and citation frequency. However, the use of agentic technology like Consensus has exposed obstacles in process visibility and control. Participants find it difficult to understand how the proxy retrieves papers, and are frustrated by the inability to intervene in intermediate steps such as modifying the query or filtering fragments without restarting the entire process. Moreover, their analysis indicates that this is a progressive process: starting with broad queries and then moving on to in-depth studies of specific subfields. This dynamic shift is not supported by rigid one-time question-and-answer tools, which treat queries as isolated events.

This study identified three key deficiencies in the current automated workflow: 

\begin{itemize}
\item \textbf{D1:} Users struggle to quickly filter papers or verify false positives of the papers' content and their directions.
\item \textbf{D2: }Due to the "black box" nature of automation, the agent prevents researchers from diagnosing or correcting intermediate errors or correcting the search logic.
\item \textbf{D3:} Current tools failed to support the natural evolution of research questions, forcing users to restart queries rather than pivoting from broad exploration to deep analysis.
\end{itemize}

Based on these observations, we developed a design solution aimed at bridging the gap between AI automation and human research needs:

\begin{itemize}
\item \textbf{R1 (Addressing D1):} Visualize the correlation indicators to assist early researchers in efficiently eliminating unsuitable papers.
\item \textbf{R2 (Addressing D1):} Ensure explicit traceability, and directly link each generated insight to its specific source, so that it can be immediately verified by humans.
\item \textbf{R3 (Addressing D2):} Expose the underlying AI logic as a visible workflow of nodes to demystify the researching process.
\item \textbf{R4 (Addressing D2):} Allow users to intervene, edit, or rerun specific intermediate nodes within the workflow without restarting the entire process.
\item \textbf{R5 (Addressing D3):} Enable users to adjust their research direction, allowing them to smoothly transition from extensive investigation and research to in-depth studies of the specific subfields identified in the previous steps.
\end{itemize}

\section{System Design}
Based on the requirements (R1-R5) identified, we developed AwesomeLit to bridge the gap between AI automation and manual verification. To power this workflow, we utilize OpenAI's \textit{gpt-5-mini}~\cite{Ope25} for all LLM-based requests, as it balances advanced reasoning capabilities with lower latency compared to larger reasoning models. For the paper resource, \textit{arXiv} API was used to get the relevant papers. Still, the approach can be configured to other powerful models and sources. As shown in Figure~\ref{fig:teaser}, the system integrates three novel features designed to facilitate steerable literature research.

\subsection {Transparent Workflow}
To bridge the gap between human intent and AI execution, the \textbf{Transparent Workflow} view introduces a novel 'process-intervention' paradigm. After users send their initial vague interest, it visualizes the otherwise opaque and uncontrollable agent working process as a directed node-link flow, demystifying the logic behind the agent's operations~\textbf{[R3]}. Each node represents a distinct functional stage in the pipeline, "Search", "Review" and "Synthesis" with execution status to provide real-time feedback. Unlike traditional static workflows, the pipeline automatically pauses after executing each node in default, requiring explicit user approval to proceed. This forces a "check-point" where the Inspector Panel displays the intermediate results for review, for instance, tracing generated chunks back to source papers~\textbf{[R2]}. Users can intervene by editing the node's output (for example, refining keywords) or rerunning the step, and only upon their confirmation does the system execute the subsequent node, ensuring granular control over the entire generation process~\textbf{[R4]}.

\subsection {Query Exploring Tree}
The \textbf{Query Exploration Tree} manages the evolution nature of the research process by visualizing it as a hierarchical tree structure. Each node represents a possible pipeline. The explored nodes will be highlighted in yellow, while the system actively proposes "possible directions" as branch nodes based on the retrieved content. This topological structure supports the non-linear workflow of literature reviews, enabling users to seamlessly transition from a broad topic to a specific subfield, or to review a previous searching state without losing context \textbf{[R5]}.

To assist decision-making during these transitions, this view incorporates two quantitative indicators: \textit{semantic offset} and \textit{semantic delta}. \textit{Semantic offset} is presented as a percentage on the connecting edge, quantifying topic deviation. Users interpret this value contextually: a high offset signals a significant pivot to a new subfield, suitable for breadth-first exploration; conversely, a low offset suggests incremental refinement, ideal for depth-first analysis. At the same time, \textit{semantic delta} clearly shows the newly added or deleted keywords that define this transition (for instance, "+ benchmark, -interpretability").

\subsection {Semantic Similarity View}
Uniquely extending the Query Exploring Tree's structural logic, the \textbf{Semantic Similarity View} serves as its evidence transforming abstract branching decisions into verifiable data distributions. We utilize OpenAI's \textit{text-embedding-3-small} model~\cite{Ope24} to generate high-dimensional embeddings for each paper's abstract, which are then projected onto a 2D plane using the Uniform Manifold Approximation and Projection (UMAP) algorithm~\cite{SWB*25}. In this scatterplot layout, spatial proximity encodes semantic similarity—papers clustered closer to the query centroid are contextually more relevant. Each paper is represented as a glyph, color-coded to reflect user interaction status: green indicates user-acceptance, red denotes explicit rejection, and blue signifies a neutral state awaiting agent assessment. Hovering over a glyph reveals a detail card containing key metadata (for example., URL, publication year, authors), enabling researchers to efficiently identify high-quality clusters and filter out irrelevant work based on visible metrics~\textbf{[R1]}. Through interactive linking with a tree selector, users can selectively highlight specific iterations, enabling them to distinguish between past search results and the current focus.
\section{Evaluation}
For evaluation we conducted a mixed-methods user study with seven target users (final-year computer science students working on final dissertations) to evaluate AwesomeLit's effectiveness in supporting hypothesis generation. 

\begin{figure}[htb]
  \centering
  \includegraphics[width=\linewidth]{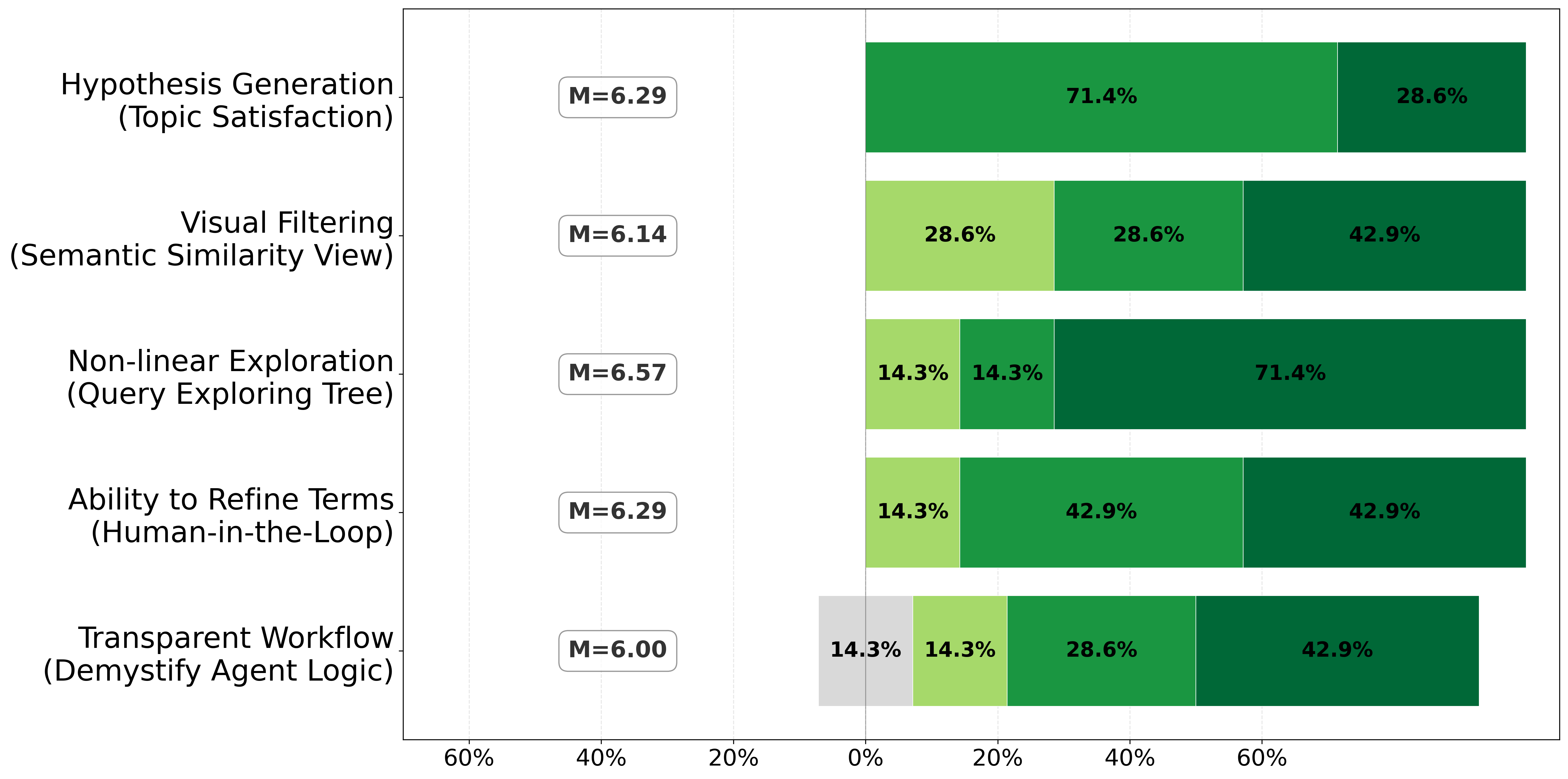}
  \caption{\label{fig:likert_results}
           Likert Chart of Participant Feedback on Usability.}
\end{figure}

Each 90-minute session for them comprised three phases: a 30 min free exploration for familiarization the 30 min targeted task session, where participants identified a novel research direction within the broad domain of "Visualization for AI", and a 30 min interview to gather qualitative feedback. We observed and recorded their working process, interaction logs, and questionnaires using a 7-point Likert scale (1=Strongly Disagree, 7=Strongly Agree) as shown in Figure~\ref{fig:likert_results}. The quantitative data was aggregated to calculate mean and for qualitative data transcripts were coded to identify recurring patterns.

The Transparent Workflow was rated highly for its ability to demystify agent logic (M=6.00). Most participants (6/7) agreed that the visual separation of "Search", "Review", and "Synthesis" nodes helped them "clearly distinguish processing stages" \textbf{[R3]}. This transparency fostered trust, enabling users to confidently verify generated insights against source papers via the "check-point" mechanism \textbf{[R2]}.

The core task required participants to generate hypothesis. As the system organized the agent pipeline in the \textbf{Query Exploring Tree}, participants expressed appreciation for the comprehensibility of this process; one noted that seeing "keywords expansion" connected to "search" eliminated the mystery surrounding the logic \textbf{[R5]}. However, initial keywords introduced overly broad terms like "XAI". The system's human-in-the-loop features proved critical. Participants utilized the breakpoint mechanism to intervene. For instance, P4, interested in interface design rather than algorithms, used the Inspector Panel to select "Interface Design" related papers steering the search toward interface focused results \textbf{[R4]}. Survey results confirmed this utility, with 6/7 participants rating their ability to refine search terms at 6 or higher.

During the session, the \textbf{Query Exploring Tree} acted as a visual scaffold for non-linear sensemaking. Unlike linear search tools, it encouraged participants to pivot based on emerging interests (M=6.57). We observed clear divergence in exploration paths: P3 investigated "Evaluation Benchmarks" direction while P6 explored "Saliency Maps". Participants reflected that the tree structure helped them "remind relationships between topics" reducing the cognitive burden of tracking complex histories compared to linear chats during the interview. Concurrently, the \textbf{Semantic Similarity View} enabled efficient visual filtering of relevant literature (M=6.14) \textbf{[R1]}.

All participants successfully narrowed down the broad topic into sub-topic from "Visual Analytics for Bias Detection" to "Interactive Steering for Generative Editing" and raised their satisfied hypotheses. The interview results highlighted that the system made exploration more reasonable by organizing the relationship between each pipeline, although minor suggestions for improved color coding in \textbf{Semantic Similarity View} were noted in the written feedback.

\section{Conclusion \& Future Work}
In a nutshell, AwesomeLit effectively addresses key deficiencies by providing a transparent interface for rapid verification, enabling granular intervention in the agent’s workflow, and guiding the inquiry process from broad exploration to specific analysis. However, it still relies heavily on manual guidance. Future improvements could include adaptive user profiling, where the system learns from user interactions to recommend relevant papers and directions. 
In conclusion, AwesomeLit presents a novel approach to Human-AI collaborative research by integrating transparent agent workflows with semantic visualizations. Enabling researchers to visually track their exploration path, intervene in the agent's decisions, and steer their topics' evolution, the tool supports a structured and verifiable workflow for early-stage literature reviews. Its key contribution lies in its ability to combine the agentic research process, semantic relevance, and topic evolution—into a unified, steerable scaffolding tool tailored for academic discovery.

\newpage
\bibliographystyle{eg-alpha-doi} 
\bibliography{egbibsample}     
\end{document}